# PROTEINS AS BIOELECTRONIC MATERIALS: ELECTRON TRANSPORT THROUGH SOLID-STATE, PROTEIN MONOLAYER JUNCTIONS.

Izhar Ron[a], Lior Sepunaro[a], Stella Izhakov[a], Noga Friedman[b], Israel Pecht[c,*], Mordechai Sheves[b,*] and David Cahen [a,*]

Departments of [a]Materials and Interfaces, [b]Organic Chemistry and [c]Immunology

Weizmann Institute of Science, POB 26, Rehovot 76100, Israel

## Abstract

Electron transfer (ET) through proteins, a fundamental element of many biochemical reactions, has been studied intensively in solution. We report the results of electron transport (ETp) measurements across proteins, sandwiched between two solid electrodes with a long-range goal of understanding in how far protein properties are expressed (and can be utilized) in such a configuration. While most such studies to date were conducted with one or just a few molecules in the junction, we present the high yield, reproducible preparation of large area monolayer junctions of proteins from three different families: Azurin (Az), a blue-copper ET protein, Bacteriorhodopsin (bR), a membrane protein-chromophore complex with a proton pumping function, and Bovine Serum Albumin (BSA). Surprisingly, the current-voltage (I-V) measurements on such junctions, which are highly reproducible, show relatively minor differences between Az and bR, even though the latter lacks a known ET function. ETp across both Az and bR is much more efficient than across BSA, but also for the latter the currents are still high, and the decay coefficients too low to be consistent with coherent tunneling. Rather, inelastic hopping is proposed to dominate ETp in these junctions. Other features such as asymmetrical I-V curves and distinct behavior of different proteins can be viewed as molecular signatures in the solid-state conductance.

## Introduction

**E**lectron Transfer (ET) is one of the most fundamental processes in biological systems(1), crucial for different biological energy conversion processes from respiration to photosynthesis, and prominent in diverse metabolic cycles. ET reactions are performed by a range of proteins (with specific components that evolved for that purpose) in which electron tunneling occurs over long distances (2). While these reactions proceed in the proteins, when present in their natural, usually aqueous environment, there have been attempts to explore the electronic conductance of various proteins in both wet electrochemical and solid-state configurations. Such efforts mainly used scanning probe microscopy (Scanning Tunneling Microscope, STM, and Conductive-probe Atomic Force Microscope, CP-AFM) (3-12). Attempts were also made to integrate proteins into solid-state devices(13-15). In most of these studies, device design was dictated by a bio-mimetic approach, namely, proteins were expected to conduct current in such metal-protein-metal electrical junctions through pathways, similar to those, known to dominate the ET process in solution. According to this notion, proteins without redox activity would be expected to behave merely as insulators in such electrical junctions. However, in earlier work of ours we found that even a protein such as Bacteriorhodopsin (bR), which has an activity known to be different from ET, namely proton pumping, is able to mediate electron transport (ETp) efficiently, when integrated into a metal-protein-metal solid-state junction. In that configuration it was shown to pass currents that are higher than can be predicted for a protein of its size (16). Therefore, it is clear that a first step towards designing protein-inspired devices is to try to understand *which factors dominate the electronic conductance of proteins in solid-state junctions*.

To pursue these issues, we set out to design the preparation of solid-state, protein monolayer junctions, to carry out reproducible ETp measurements that will allow comparative analysis of the results. The strategy for making high-quality monolayer junctions is to use, for each protein, as similar a chemical modification of a conductive substrate as possible and to allow self-assembly of the proteins on the modified surfaces. An additional requirement is to use a non-destructive method for making the top electrical contact to the soft biological monolayer. To make this study a comparative one, we chose three functionally different protein systems: 1) Azurin (Az), a bacterial type I blue Copper protein. Az is a small soluble protein, serving as an electron carrier (17). 2) Bacteriorhodopsin (bR), a membrane protein-chromophore complex that functions as a light-induced proton pump, i.e., an electro-active function, in the halophilic archea, *Halobacterium Salinarum* (18). 3) Bovine Serum Albumin (BSA), a plasma protein known to bind and transport a range of hydrophilic molecules and is readily adsorbed to surfaces. BSA has no known electro-active role in its action. As a non-protein reference that can be measured under the same conditions, we use a monolayer of long (18 carbons) saturated organic molecules, which is expected to behave as a molecular tunneling barrier.

## Results

**Protein monolayer preparation**

The preparation of solid-state, protein monolayer junctions is a three-step procedure. First a thin silicon oxide ($SiO_x$) layer is grown from an oxidizing solution on an etched surface of highly doped (degenerate) Si. We chose Si as substrate because it provides a highly reproducible flat surface, and by using highly doped p-Si it serves as an electrode with minimal semiconductor-related effects. Next, a short bi-functional linker molecule is used to form a monolayer, sufficiently dense to control the tunneling through it (and not through defects, which would be the case if some $SiO_x$ surface were inadvertently exposed to the top electrode), and to allow the proteins to cover the surface completely. Finally, proteins are adsorbed from aqueous solutions/suspensions by immersing the chemically modified substrates in the protein solutions for times depending on the type of protein-linker bond that is used for the formation of the monolayer. A schematic representation of this process is shown in Figure 1. The long organic monolayer was assembled from Octadecyl-trimethoxysilane molecules (OTMS) on top of the oxidized Si.

Three types of linker molecules (all of similar length) were used, viz., a propylsilane chain terminated either by an Amine, a Bromide or Thiol group, for interaction with the proteins.

Az was adsorbed to either Br- or SH- terminated surfaces, yielding identical results in terms of thickness and morphology of the monolayer, as well as in electrical transport characteristics. Az has a disulfide bridge at its surface (on the opposite end of the Cu site) which is commonly used as an anchoring unit to Au surfaces. Az molecules were covalently linked to the thiol-terminated surface presumably by means of S-S bond formation or by substitution of the Br group, in the case of Br-terminated surface.

Bacteriorhodopsin was adsorbed electrostatically on amine-terminated substrates, because its vesicles are negatively charged on both sides, with the cytoplasmic side being more negative (19). The mixed protein-lipid monolayer formed from native bR protein and lipids (20), which underwent vesicle fusion upon adsorption.

BSA was adsorbed on either of the three types of substrates by immersing the substrates in its solution. Because BSA is known to adhere to most surfaces, we cannot conclude at this point which group of the protein actually anchors it to each surface, but we assume that the BSA aligns with its major axis parallel to the surface (see below).

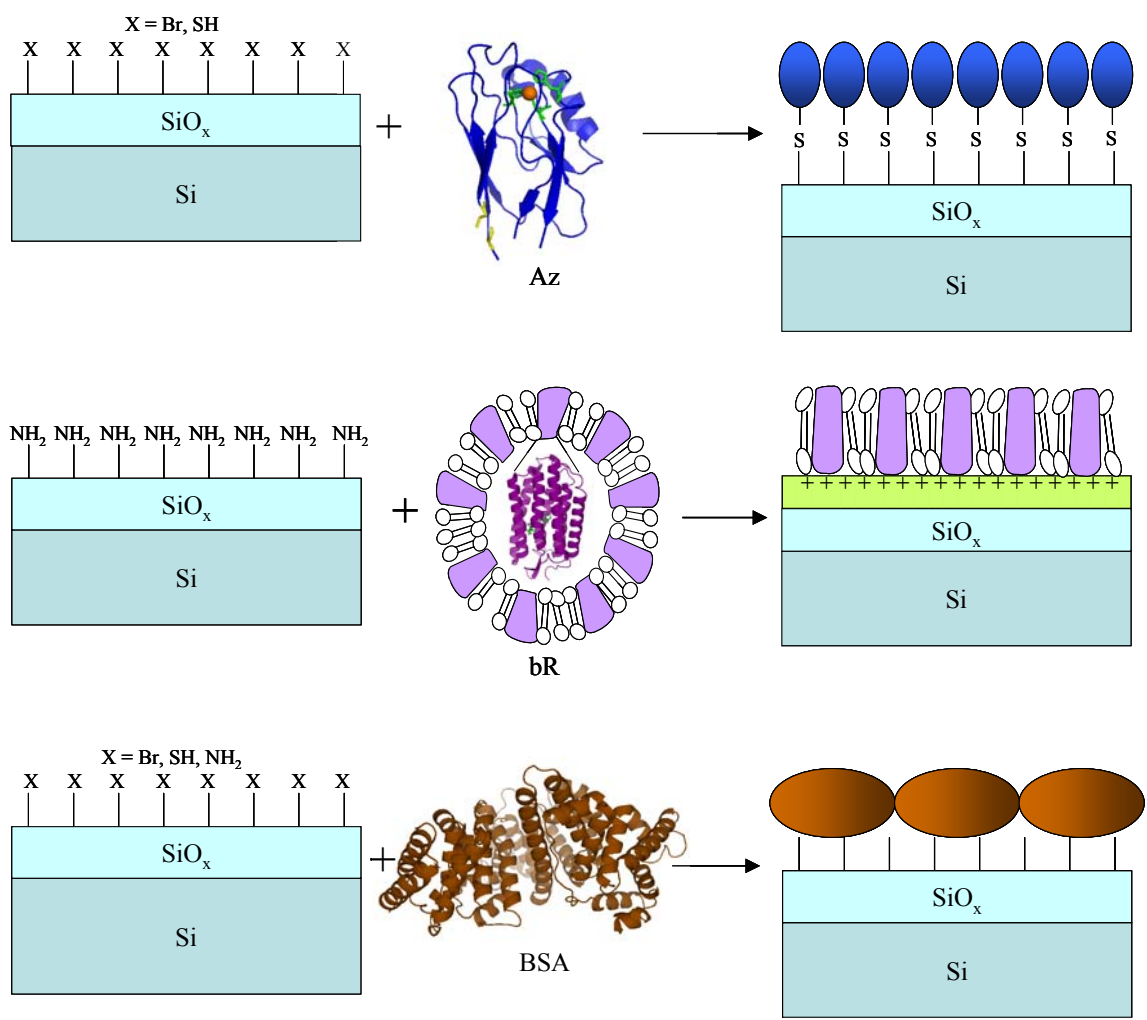

Figure 1: Schematic representation of the protein monolayers assembly process (coordinates were taken from Protein Data Bank, PDB, code 1AZU for Az, code 1R2N for bR; BSA model was obtained from ModBase).

**Monolayer surface characterization**

Ellipsometry measurements were carried out on all samples after each preparation step, i.e., oxide growth, silane monolayer formation and protein monolayer formation. The thickness, deduced from the ellipsometry results, together with reported crystallographic dimensions of the proteins and the results of surface roughness analysis by Atomic Force Microscopy (AFM) imaging are summarized in Table 1. The ellipsometry results help to evaluate if a dense, homogenous monolayer was obtained, if used in combination with AFM characterization, as will be explained later. The Si oxide layer was 11-12 Å thick, and the silane monolayer 6-7 Å, in all samples. Data for the organic and protein layers were analyzed with a Cauchy model.

Az monolayer thickness was deduced from ellipsometry to be 13-16 Å. This value agrees with reported values of Az monolayers, derived from ellipsometry and X-ray Photoemission Spectroscopy (XPS). Those values were supported by a theoretical model that simulated the thickness of a layer of globular shapes with the size of the Az molecules, taking into account the voids that exist between them, if adsorbed on a planar surface(21).

bR monolayer thickness was calculated, from ellipsometry data, to be 75-80 Å, significantly higher than the known thickness (~ 50 Å) of the purple membrane (which is also the height of the bR protein embedded therein). This difference can probably be attributed to the presence a conjugated system (the retinal chromophore) bound to the bR protein. Such a conjugated system affects the (optical) ellipsometry measurement analysis, which uses the refractive index of the layer. As will be shown later, AFM topography rules out the possibility of over-layer formation.

BSA thickness was 12-14 Å on the $NH_2$-terminated surface, 16-20 Å on the Br-terminated surface and 20-22 Å on the SH-terminated surface. BSA dimensions, according to models, are 40•40•140 Å, which represents an ellipsoidal shaped (22). The measured thickness, within the above-mentioned model of a porous macromolecule used for Az, and our AFM characterization, allows us to describe the BSA as a barrel-shaped structure, aligned on the surface with its minor axes perpendicular to the surface(23). The variation in BSA monolayer thickness on substrates with different end groups may result from anchoring by different protein groups, which may lead to slight variations in the orientation in which it is adsorbed. OTMS thickness by ellipsometry was 22-24Å.

Table 1: Surface characterization parameters of the monolayers.

| | Ellipsometry-derived thickness (Å) | Crystallographic/ theoretical size (normal to surface) (Å) | RMS roughness, from AFM (Å) |
|---|---|---|---|
| $SiO_x$ | 11-12 | -- | 2 |
| Organo-Silane ($NH_2$, Br, SH end groups) | 6-7 | 7 | 2.5 |
| OTMS | 22-25 | 24 | -- |
| bR | 75-80 | 50 | -- |
| Az on SH/ Br | 13-16 | 36 | 4-4.5 |
| BSA on $NH_2$/ SH/ Br | 12-14/ 20-22/ 16-20 | 40 | 5.5-6 |

AFM, performed in the Tapping Mode, which is ideal for imaging soft, solid-supported layers, served to examine the morphology of the protein monolayers. The results also complemented ellipsometry. From AFM imaging, two main features can be deduced. First, the lateral and vertical dimensions of the adsorbed species can be obtained from the image itself and from height profiles, respectively. The second approach, useful in cases where the proteins are too closely packed to evaluate their full height by scanning probe (as is the case of Az and of BSA, see below), is based on the RMS roughness of the imaged surface. Figure 2 shows typical AFM height images of the three types of monolayers.

In the Az monolayer, small globular features, covering the surface, are observed (Figure 2, left). Such features are not observed on the silane monolayer surfaces (Figure 3). The apparent lateral dimension of one globular particle in the AFM image is 170-200 Å in diameter, which corresponds to an actual size of ~ 36-50 Å, taking into account tip-sample convolution (24). The RMS roughness of this surface is 4-4.5 Å, which is only slightly higher than that of the silane-modified $SiO_x$ surface (2.5 Å) and of the bare $SiO_x$ surface (2 Å). This indicates that the Az molecules are densely packed, and, therefore, the full height of the protein (36 Å in the proposed orientation) cannot be measured. Combined with the ellipsometry data we conclude that the Az monolayer is homogeneous and mostly defect-free.

The bR monolayer is made up of similarly sized fused vesicles that are closely spaced on the substrate (Figure 2, middle). The height profile indicates that most vesicles are 50-55 Å high, in keeping with the crystallographic data. Surface roughness analysis is, therefore, not relevant in this case. This observation supports the above-mentioned assumption that the higher ellipsometry-derived height value (75-80 Å) is related to the nature of this protein rather than to its structure.

The BSA monolayer exhibited elongated features with apparent longer dimension measured to be ~300-360 Å (Figure 2, left), which translates to 110-160 Å after correcting for tip-sample convolution, in keeping with model estimations of the protein's long dimension. The full height could not be measured here as well, probably for the same reason as for the case of Az. The RMS roughness was 5.5-6 Å, higher than that of the Az monolayer. This is reasonable if one takes into account the internal changes of height within a single BSA molecule, which are larger than for Az. In this case, the AFM image itself provides a good morphological picture of the arrangement of the BSA molecules on the surface, and again, in combination with the ellipsometry data, indicates a homogenous layer with high coverage of the surface.

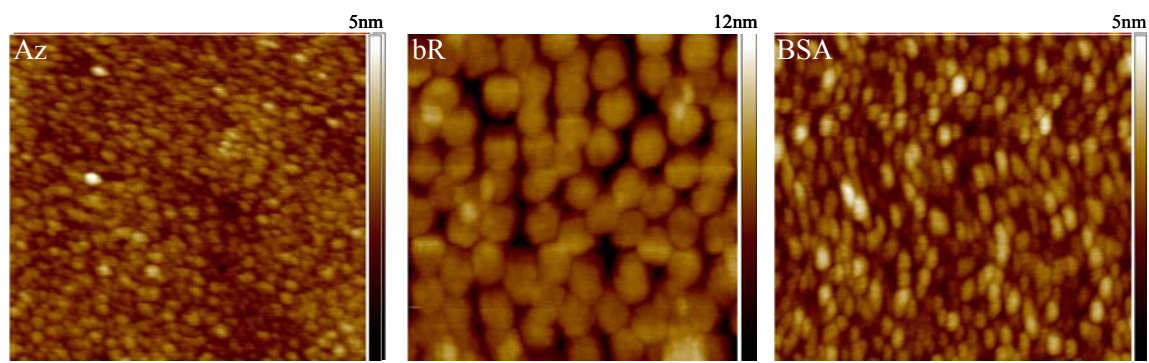

Figure 2: AFM images (500nm x 500nm) of Az (left), bR (middle) and BSA (right) monolayers.

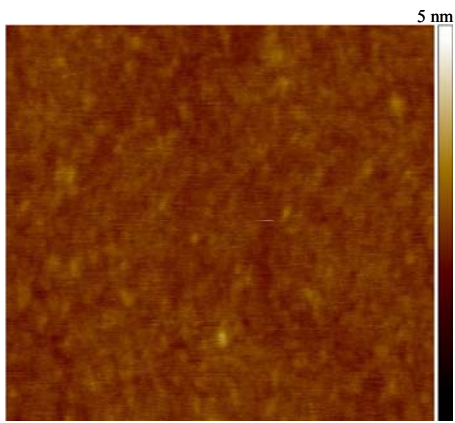

Figure 3: AFM image (500nm x 500nm) of APTMS ($NH_2$-terminated silane) monolayer.

**Electrical transport measurements**

ETp measurements were carried out with either a hanging Mercury drop (25) or a 60 nm thick "ready-made" Au pad, deposited from water onto the monolayer (LOFO) (26), as second electrode/contact. We measured the samples using the two methods separately, so as to eliminate possible effects of a specific metal and contacting method on the junctions' transport characteristics. Because similar results were obtained with both methods we will refer to the results obtained with the Hg drop top contact only, simply because practically it is easier to collect data with this method than with the Au one. Still, ETp results for Az and bR, obtained with Au and Hg top contacts are shown in Figure 4 (27).

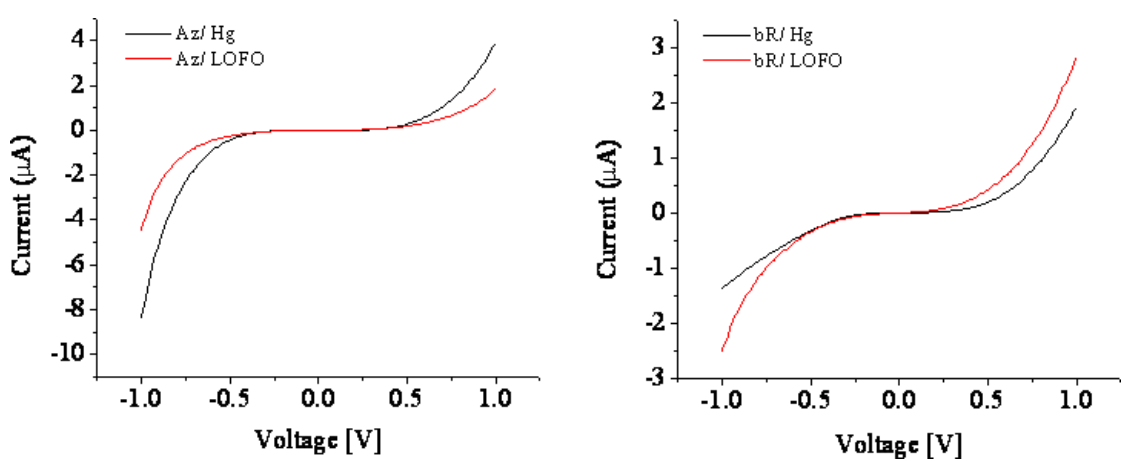

Figure 4: I-V characteristics of Az (left) and bR (right) with the two top contacting methods (soft deposition of Au film, LOFO, and hanging mercury drop, Hg).

Figure 5 shows I-V curves of the three different linker monolayers. The current magnitudes extracted from these curves will later be used as the initial current, entering the protein monolayers that assembled on top of these linker molecules. It is evident that for all three linkers, the current at ±1 V bias voltage is on the order of mA. The bare $SiO_x$ surface behaves more as an Ohmic junction, with current magnitudes of ~35 mA at ±1V (Figure 6).

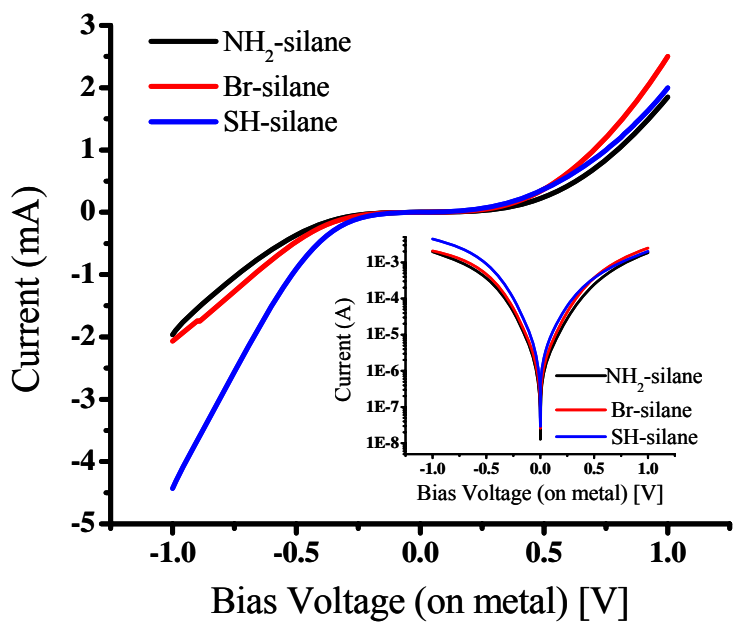

Figure 5: I-V curves of the organosilane monolayers. Inset: Semi-Log plot of the curves

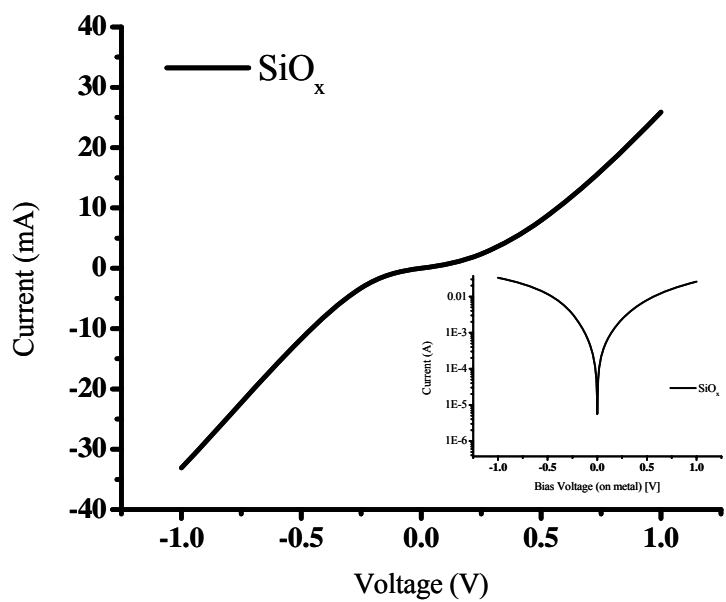

Figure 6: I-V characteristics of the bare $SiO_x$ substrate (12Å oxide layer). Inset: Semi-Log plot of the curve

Figure 7 presents I-V curves of Az on the SH-terminated substrate, bR on NH$_2$-terminated substrate, BSA on NH$_2$-terminated substrate (I-V curves of Az and BSA on Br-terminated substrates are shown in Figure 8) and OTMS on SiO$_x$.

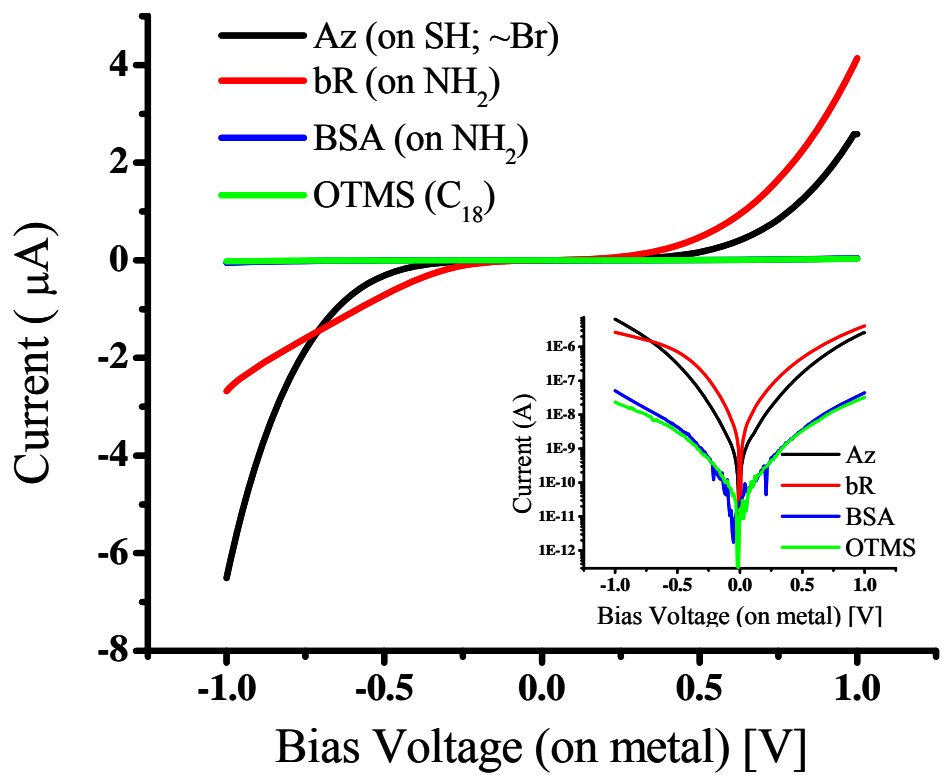

Figure 7: I-V curves of the three protein monolayers and the OTMS organic monolayer. Inset: Semi-Log plot of the curves

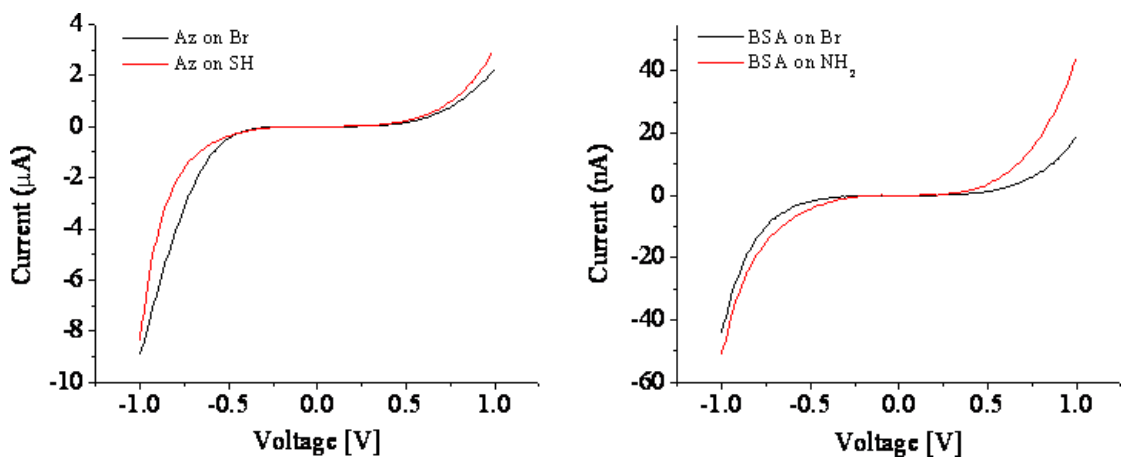

Figure 8: I-V curves of Az on Br-terminated and SH-terminated substrates (left) and of BSA on Br-terminated and NH$_2$-terminated substrates (right).

Three trends are immediately apparent from these curves. First, currents through bR and Az monolayers are of the same magnitude; this is surprising, considering the size of bR compared to Az, and more important, bR's lack of known ET functionality, as opposed to Az. Second, currents through Az and bR are on the order of µA at ± 1 V, while currents through BSA and OTMS are on the order of tens of nA (28). The third issue concerns the shape of the I-V curves. The curves for bR show asymmetry towards positive bias while those for Az show asymmetry towards negative bias. The asymmetry ratio in both cases is ~2. The I-V curve for just the underlying substrate relevant for bR ($NH_2$-terminated substrate) was almost symmetric, while asymmetry of Az samples was observed with the two different linkers (Br- and SH-terminated substrates), even though these two substrates alone showed opposite asymmetries (positive with Br and negative with SH). We, therefore, suggest that the observed asymmetry reflects properties of the proteins in the junctions. The following reasons strongly indicate that ETp occurs through the proteins and not through other routes in the junction:

- the pronounced difference between current magnitude observed with Az and bR, and that with BSA,
- the difference in ETp between Az and bR on the one hand and OTMS (a much shorter barrier), on the other hand
- the similarity between OTMS and BSA, notwithstanding the large difference in size between them
- the rough similarity between liquid Hg and "ready-made" Au pad contacts (29)
- and the much higher currents measured through the linker molecules only.

On this basis we will now try to provide the reasons for the observed results.

**ETp through protein junctions**

As already noted, currents through Az and bR are of comparable magnitudes. Both monolayers attenuate the currents that could pass through the underlying linker monolayers, by about three orders of magnitude. This result is surprising, as one could naively expect that only an ET protein such as Az will allow efficient ETp in such a junction, while bR should, according to this approach, function as a mere insulator. Already in our earlier work we noted the remarkably efficient ETp through bR (20), but we could not put it in perspective. This is now possible by comparing bR ETp with that of two other proteins, measured under the same conditions. To quantify the ETp differences between the protein systems we will first refer to the protein monolayer junctions as simple tunneling junctions, an approach which implies two strongly simplifying assumptions:

1. *Current can be described by a simple mono-exponential decay, $I = I_0 \cdot exp(-\beta \cdot l)$*. We can then extract a value for *β*, the inverse range or tunneling decay parameter. To this end we use the currents, measured before assembly of the protein monolayers, as $I_0$ values and the protein's crystallographic size, *l*, as tunneling barrier width. *β* is an intrinsic property of the junction, a parameter into which many processes are lumped together. Extracting a value for it from a mono-exponential term is valid only for coherent tunneling, the validity of which will be examined later considering an alternative scenario.

2. Even though the geometrical area is the same for all junctions, the number of molecules per unit area is different, because the three described proteins are not of the same size. Therefore, the number of molecules, contacted by the top electrode and participating in the ETp process is not the same for all protein junctions. However, because, at this point, we do not know which path(s) the electrons follow in the proteins, we shall refer to these junctions as comprising conduction pathways where a single protein may contain several such pathways (30). Therefore, *we consider current through the geometrical surface area of the contact*, rather than through a single protein, assuming that conduction pathways are closer spaced than the sizes of the proteins used here.

From this analyses (Table 2), for the OTMS monolayer junction, $\beta = 0.58$ Å$^{-1}$ was extracted, which falls within the range of decay constants proposed for tunneling through non-conjugated organic molecules (31). However, all three protein-containing junctions yield *β* values that are significantly lower, with bR yielding an even lower value (0.12Å$^{-1}$) than Az (0.18Å$^{-1}$) and BSA giving the highest ETp decay constant among the three proteins (0.27Å$^{-1}$). How can we understand that all three proteins, regardless of their functions and possible unique features give such *low, <0.3Å$^{-1}$, β* values, is a question we will return to later.

Table 2: Electrical Transport characteristics of the monolayers
(In parentheses- the substrate from which $I_0$ was derived).

|      | $I_0$ (mA) at ± 1 V | $l$ (Å) | $I$ (nA) | $\beta$ (Å$^{-1}$) |
|------|---------------------|---------|----------|--------------------|
| OTMS | 26 (SiO$_x$)        | 24      | 23       | 0.58               |
| Az   | - 4.43 (SH)         | 36      | -6,500   | 0.18               |
| bR   | 1.85 (NH$_2$)       | 50      | 4,140    | 0.12               |
| BSA  | 1.85 (NH$_2$)       | 40      | 44       | 0.27               |

## Discussion

An important achievement in this study is that the protein monolayers that we prepared are homogeneously dense and suitable for electrical transport measurements. Indeed, using top electrodes that do not damage the protein monolayers, ETp measurements are very reproducible. For all the described monolayers, dozens of junctions could be measured on samples from different monolayer preparations, without the occurrence of short circuit currents (currents that are dominated by pinholes or imperfections in the protein monolayers, see electrical measurement part in materials and methods). Therefore, this system can be analyzed reliably trying to clarify how the different proteins conduct current.

By extracting current values observed at a given voltage, and by comparing them with those measured through the same junctions *without* the protein layer, we have deduced the average tunneling decay coefficients of the protein junctions. These values were not only much smaller than those derived from calculations(32) and from (mostly optical) solution ET measurements(33, 34), but also much smaller than values that were assigned to proteins or other species, referred to as insulators in solid-state junctions (9, 31).We therefore suggest that for ETp, proteins cannot be viewed as providing a simple molecular tunneling barrier, in contrast to what works for saturated hydrocarbon chains. Still, the two types of proteins that have charge transport as their biological function (bR and Az) are found to be significantly better conductors than the electro-inactive protein (BSA).

Two main models have been proposed to explain ETp in <u>molecular</u> junctions: *Superexchange*, that treats ETp as a coherent tunneling process, mediated by virtual states of the molecular bridge (protein in our case), but electrons do not populate the bridge levels (35). This process is characterized by a relatively strong distance decay and weak temperature dependence. The second common model is that of inelastic charge *hopping*, where electrons (or holes) travel through the bridge by short tunneling steps from one hopping site to the next, in a process that is only weakly dependent on distance(36). Possible combinations of the two processes have also been proposed(37).

The decay constants that we observe for ETp through protein, are much lower than those that describe tunneling through proteins in solution, where proteins were treated as featureless barriers (32, 38), which may indicate that a different mechanism is operating in our case. For ET in solution a marked dependence on the distance between donor and acceptor was observed. In the present studies we measured ETp across the assumed *full length* of the proteins. Therefore, a system that is similar but spans a range of lengths may allow the resolution of the two above-mentioned mechanisms. Demonstration of ET through such a range of distances has been found in ET studies through peptides, with lengths ranging from a few to few dozens of Å.

Studies on peptides are relevant here, because it has been proposed that the structure of proteins might have evolved to support long range ET(39). Indeed, from a chemical point of view, charged amino acid residues, hydrogen-bonded networks and amide bonds that are present along the polypeptide chain, can be considered as excellent candidates for hopping sites(40). Experimental studies of ET through peptides with varying lengths (41, 42), supported by theoretical work (43), showed that for short peptides, higher $\beta$ values are extracted (~1.4 Å$^{-1}$) and, for the same type of peptides, from a certain length (~20 Å) upwards, a transition into much lower decay ($\beta$) has been observed (~0.2 Å$^{-1}$). This suggests that only for the longer peptides hopping becomes favorable over superexchange. We can understand this by considering that over longer distances direct tunneling becomes less probable and at the same time hydrogen-bonded networks span larger parts of the bridge. Thus, the number of hopping sites increases, favoring inelastic charge hopping (44). While an OTMS monolayer is long enough to be compared to peptides, here hopping is unfavorable because the above-mentioned features of proteins that support hopping are absent. Therefore, a stronger decay is observed, suggesting that for such molecules there is no alternative to superexchange.

We propose that an inelastic hopping model may be appropriate for describing the behavior of our junctions, and of protein junctions in general. The advantageous ETp via Az and bR over BSA can be explained by the additional factors that may control ETp in solid state junctions, such as the redox site in Az, the retinal chromophore and/or the electrostatically screened proton channel in bR, and possibly the presence of more favorable conduction pathways in both, as part of how the design of these proteins evolved for charge transport.

Finally, we note that the observed asymmetrical transport behavior in both Az and bR junctions is uncommon in junctions dominated by coherent tunneling (45), if all the potential drops over the interfaces, which usually results in symmetrical curves. Substrate-independent asymmetries may suggest that ETp across Az and bR is structure- sensitive, so that our monolayer junctions approach can be used to study point changes in the proteins. Among these point changes are the removal as well as replacement of Cu with another metal in Az, the removal of the retinal from bR as well as its replacement with an analog, and various point mutations for studying the effect of specific amino acids; Trp for example, present in Az and bR, is one appealing candidate in this sense, considering its role in facilitating ET in proteins (46, 47).

**Conclusions**

We have described the highly reproducible preparation of solid-state, protein monolayer junctions, with three distinct types of proteins, and demonstrated high-yield electrical transport measurements on these junctions. Our results clearly show that even a system that does not handle ET naturally (bR) can facilitate current flow in the solid-state. The charge transporting proteins (Az and bR) seem to provide more efficient ETp than a non-charge carrying protein (BSA), which in turn still conducts better than a simple saturated organic molecule (OTMS). The ETp across the proteins cannot be readily interpreted by the simple coherent tunneling behavior that is commonly used for organic monolayer junctions. Sequential inelastic transport seems more appropriate for rationalizing our results at this stage, in keeping with previous interpretations of ET through peptide junctions. Our preparation and measurement methods can serve as a general platform for studying ETp across proteins in a solid-state configuration. The remarkable current densities (on the order of mA/cm$^2$) that were measured indicate that proteins should not be viewed as insulators, and electronic current dependence on one of the protein's functionalities may be pursued under these conditions, as a basis for biomolecule-based devices.

**Materials and Methods**

*Monolayer preparation -* Highly-doped (<0.005 ohm•cm) p-type Silicon wafers <100> were cleaned by bath- sonication in ethyl acetate/ acetone/ ethanol (2 min. in each), followed by 30 min. of piranha treatment (7: 3 V: V of $H_2SO_4$: $H_2O_2$) at 80º C. The wafers were then thoroughly rinsed in Milli-Q (18 MΩ) water, dipped in 2% HF solution for 1 min. in order to etch the Si surface (leaving a Si-H surface) and put in fresh piranha for 25 min. for controlled growth of the oxide layer. After this step, the wafers were thoroughly rinsed in water and dried under a Nitrogen stream. The resulting SiO$_x$ layers served as a substrate for preparation of three different organo-silane layers:

1) 3-Aminopropyltrimethoxysilane (3-APTMS, NH$_2$-terminated linker, ALDRICH) monolayer was prepared by immersing the SiO$_x$ substrate in 10% V: V 3-APTMS in methanol for 3 hr., followed by 3 min. bath-sonication in methanol and rinsing in water.

2) 3-Bromopropyl trichlorosilane (3-BPTCS, Br-terminated linker, ALDRICH) monolayers were prepared by immersing the SiO$_x$ substrate in 10mM 3-BPTCS in bicyclohexyl (BCH) for 30 sec., followed by 2 min. bath-sonication in toluene and rinsing in ethanol.

3) 3-Mercaptopropyltrimethoxysilane (3-MPTMS, SH-terminated linker, FLUKA) monolayers were prepared by immersing the SiO$_x$ substrate in 10 mM 3-MPTMS in bicyclohexyl (in the presence of 5mM DTT), overnight, followed by 2 min. bath-sonication in acetone and rinsing in ethanol.

*Proteins* – Bacteriorhodopsin: a suspension of purple membrane fragments containing wild-type bR was prepared by a standard method(48). Membrane vesicles were prepared by following the procedure of Kouyama *et al.*(49).

Azurin was isolated from Pseudomonas aeruginosa by the method of Ambler and Wynn (50).

A280/A625 values of the isolated and purified protein were 2.0.

BSA was prepared by dissolving Fraction V powder (SIGMA) in buffer solution.

***Protein Monolayers*** - Bacteriorhodopsin monolayers were prepared by immersing the $NH_2$-terminated substrates in bR vesicle suspension for 15 min. followed by transferring the sample to water and keeping it for three hr. to allow vesicle fusion. After taking the substrates out they were gently rinsed in water and dried under a fine Nitrogen stream.

Azurin monolayers were prepared by immersing the SH- and Br-terminated substrates in 1mg/ml solution of Azurin in 50 mM Ammonium Acetate ($NH_4Ac$) buffer, pH 4.6, for three hr. followed by rinsing in clean 50mM $NH_4Ac$ buffer and finally in $H_2O$, followed by drying under a fine Nitrogen stream.

BSA monolayers were prepared by immersing the $NH_2$-, Br and SH-terminated substrates in 1mg/ml solution of BSA in 20 mM phosphate buffer, pH 7, for two hr. followed by rinsing in clean 20 mM phosphate buffer and finally in $H_2O$, followed by drying under a fine Nitrogen stream.

***Back contacts*** - before the electrical transport measurements, the back side of all samples was scratched with a diamond pen and In-Ga eutectic was applied to the back of the sample. The samples were mounted on a conducting sample holder.

***Top metal electrode deposition-*** Au pads (60 nm-thick, 0.5 mm diameter) were evaporated on clean glass slides. The pads were lifted off the glass slide by immersing them in 2% HF solution and then dipping in $H_2O$, to allow the pads to float on the water surface. Samples were dipped into the water and pulled out until a pad was deposited on the surface. After depositing several pads on each sample, the sample was left to dry overnight in ambient conditions. During the measurements these contacts were contacted by a 35 μm-wide Au wire that was attached to a W probe, mounted on a micromanipulator. The approach of the Au wire to the Au contact was monitored by an optical microscope.

Hg drop top contacts were applied by placing a Hg drop (99.9999% purity) on the monolayer, using a controlled growth hanging mercury drop (HMD) electrode apparatus (Polish Academy of Sciences, Poland). The samples were mounted on a conducting sample holder, whose position was controlled by a micromanipulator. The approach of the sample towards a fresh Hg drop was monitored by an optical microscope and CCD camera.

The geometrical contact areas of the Au and Hg contacts were 0.002 cm$^2$ and 0.002±0.0005 cm$^2$, respectively.

***Electrical measurements-*** I-V curves were measured on samples from at least three different preparations. In each preparation at least two separate samples (of each type) were measured. The Hg drop was used to contact 10 points on each sample, which was 8 mm • 12 mm in size. Before and after the measurement of each sample, and after every few spots on the sample, the Hg drop was used to measure a reference sample. *Current-Voltage (I-V)* characteristics that are shown are the average of at least 60 different junctions, on three different sample preparations for each type of protein. The standard error of this averaging is less than 10%. Short circuit junctions were observed in less than 5% of the measurements (more details can be found in supporting information). Before and after the measurement of each sample, and after every few spots on the sample, the Hg drop was used to measure a reference sample (usually a linker monolayer that corresponded to the linker type, used with the protein that was measured), in order to make sure that the drop did not interact with the proteins in the monolayer and was not affected by the measurements. If the reference measurement changed, the drop was replaced with a new one (and the last measurement was discarded). Curves were recorded by applying bias on the top metal contact and sweeping it from 0 V to -1 V, than to 1 V and back to 0 V, in order to avoid electrical load of the junction with the beginning of the voltage sweep. The curves that are shown are the -1 V to 1 V part of the voltage scan. *Current-Voltage (I-V)* characteristics that are shown are the average of at least 60 different junctions, on three different sample preparations for each type of protein. The standard error of this averaging is less than 10%. Short circuit junctions were observed in less than 5% of the measurements, i.e., only this fraction of junctions had characteristics, typical of samples without proteins, or of bare SiO$_x$, which could be interpreted as pinholes in the monolayer that allowed complete penetration of the top contact into the layers underlying the protein monolayer. Such reproducibility is encouraging, considering the large area monolayers of relatively flexible species such as proteins.

***Instruments-*** *Ellipsometry* measurements were performed with a Woollam M-2000V multiple wavelength ellipsometer at an angle of incidence of 70º.

*AFM Imaging* was performed in the Tapping Mode, using a Nanoscope V Multimode AFM (Veeco, USA) and standard Si probes for AC mode AFM (OMCL-AC240TS-W2, Olympus, Japan).

*Current-Voltage (I-V)* measurements were performed using a Keithley 6430 Sub-Femtoamp Source-Meter, with a voltage scan rate of 20mV/sec.

**Acknowledgements**

We thank Prof. Mati Fridkin (WIS) for advising on protein adsorption. We thank the NATO Science for Peace Program (DC), the Nancy and Stephen Grand Centre for Sensors and Security (DC), the Gerhard Schmidt Minerva Centre for Supramolecular Chemistry (DC), the Kimmel Centre for Nanoscale Science (DC), the Ilse Katz Centre for Materials Research (MS, DC) for partial support. MS holds the Katzir-Makineni professorial chair in chemistry and DC holds the Sylvia and Rowland Schaefer chair in Energy Research.

**References**

1. Marcus, R. A. & Sutin, N. (1985) Electron Transfers In Chemistry And Biology, *Biochimica Et Biophysica Acta* **811,** 265-322.
2. Winkler, J. R., Di Bilio, A. J., Farrow, N. A., Richards, J. H. & Gray, H. B. (1999) Electron tunneling in biological molecules, *Pure And Applied Chemistry* **71,** 1753-1764.
3. Axford, D. N. & Davis, J. J. (2007) Electron flux through apo- and holoferritin, *Nanotechnology* **18,** 145502.
4. Casuso, I., Fumagalli, L., Samitier, J., Padros, E., Reggiani, L., Akimov, V. & Gomila, G. (2007) Electron transport through supported biomembranes at the nanoscale by conductive atomic force microscopy, *Nanotechnology* **18**, 465503.
5. Delfino, I., Bonanni, B., Andolfi, L., Baldacchini, C., Bizzarri, A. R. & Cannistraro, S. (2007) Yeast cytochrome c integrated with electronic elements: a nanoscopic and spectroscopic study down to single-molecule level, *Journal Of Physics-Condensed Matter* **19,** 225009.
6. Lee, I., Lee, J. W. & Greenbaum, E. (1997) Biomolecular electronics: Vectorial arrays of photosynthetic reaction centers, *Physical Review Letters* **79,** 3294-3297.
7. Reiss, B. D., Hanson, D. K. & Firestone, M. A. (2007) Evaluation of the photosynthetic reaction center protein for potential use as a bioelectronic circuit element, *Biotechnology Progress* **23,** 985-989.

8. Stamouli, A., Frenken, J. W. M., Oosterkamp, T. H., Cogdell, R. J. & Aartsma, T. J. (2004) The electron conduction of photosynthetic protein complexes embedded in a membrane, *Febs Letters* **560,** 109-114.

9. Zhao, J. W., Davis, J. J., Sansom, M. S. P. & Hung, A. (2004) Exploring the electronic and mechanical properties of protein using conducting atomic force microscopy, *Journal Of The American Chemical Society* **126,** 5601-5609.

10. Ron, I., Friedman, N., Cahen, D. & Sheves, M. (2008) Selective Electroless Deposition of Metal Clusters on Soild-Supported Bacteriorhodopsin: Applications to Orientation Labeling and Electrical Contacts, *Small* **4,** 2271-2278.

11. Alessandrini, A., Corni, S. & Facci, P. (2006) Unravelling single metalloprotein electron transfer by scanning probe techniques, *Physical Chemistry Chemical Physics* **8,** 4383-4397.

12. Friis, E. P., Andersen, J. E. T., Kharkats, Y. I., Kuznetsov, A. M., Nichols, R. J., Zhang, J. D. & Ulstrup, J. (1999) An approach to long-range electron transfer mechanisms in metalloproteins: In situ scanning tunneling microscopy with submolecular resolution, *Proceedings Of The National Academy Of Sciences Of The United States Of America* **96,** 1379-1384.

13. Carmeli, I., Frolov, L., Carmeli, C. & Richter, S. (2007) Photovoltaic activity of photosystem I-based self-assembled monolayer, *Journal Of The American Chemical Society* **129,** 12352-+.

14. Das, R., Kiley, P. J., Segal, M., Norville, J., Yu, A. A., Wang, L. Y., Trammell, S. A., Reddick, L. E., Kumar, R., Stellacci, F., Lebedev, N., Schnur, J., Bruce, B. D., Zhang, S. G. & Baldo, M. (2004) Integration of photosynthetic protein molecular complexes in solid-state electronic devices, *Nano Letters* **4,** 1079-1083.

15. Maruccio, G., Biasco, A., Visconti, P., Bramanti, A., Pompa, P. P., Calabi, F., Cingolani, R., Rinaldi, R., Corni, S., Di Felice, R., Molinari, E., Verbeet, M. R. & Canters, G. W. (2005) Towards protein field-effect transistors: Report and model of prototype, *Advanced Materials* **17,** 816-+.

16. Jin, Y. D., Friedman, N., Sheves, M., He, T. & Cahen, D. (2006) Bacteriorhodopsin (bR) as an electronic conduction medium: Current transport through bR-containing monolayers, *Proceedings Of The National Academy Of Sciences Of The United States Of America* **103,** 8601-8606.

17. Adman, E. T. (1985) in *Topics in Molecular and Structural Biology: Metalloproteins*, ed. Harrison, P. M. (Chemie Verlag, Weinheim, Germany), pp. 1-42.

18. Oesterhelt, D. & Stoeckenius, W. (1971) Rhodopsin-like Protein from the Purple membrane of Halobacterium Halobium, *Nature (London) New Biology* **233,** 149-154.

19. He, T., Friedman, N., Cahen, D. & Sheves, M. (2005) Bacteriorhodopsin monolayers for optoelectronics: Orientation and photoelectric response on solid supports, *Advanced Materials* **17,** 1023-+.

20. Jin, Y. D., Friedman, N., Sheves, M. & Cahen, D. (2007) Bacteriorhodopsin-monolayer-based planar metal-insulator-metal junctions via biomimetic vesicle fusion: Preparation, characterization, and bio-optoelectronic characteristics, *Advanced Functional Materials* **17,** 1417-1428.

21. Schnyder, B., Kotz, R., Alliata, D. & Facci, P. (2002) Comparison of the self-chemisorption of Azurin on gold and on functionalized oxide surfaces, *Surface And Interface Analysis* **34,** 40-44.

22. Peters, J. & Peters, T. J. (1995) *All About Albumin: Biochemistry, Genetics and Medical Applications* (Academic, San Diego, CA).

23. Mori, O. & Imae, T. (1997) AFM investigation of the adsorption process of bovine serum albumin on mica, *Colloids And Surfaces B-Biointerfaces* **9,** 31-36.

24. Biasco, A., Maruccio, G., Visconti, P., Bramanti, A., Calogiuria, P., Cingolani, R. & Rinaldi, R. (2004) Self-chemisorption of azurin on functionalized oxide surfaces for the implementation of biomolecular devices, *Materials Science & Engineering C-Biomimetic And Supramolecular Systems* **24,** 563-567.

25. Salomon, A., Boecking, T., Chan, C. K., Amy, F., Girshevitz, O., Cahen, D. & Kahn, A. (2005) How do electronic carriers cross Si-bound alkyl monolayers? *Physical Review Letters* **95**.

26. Vilan, A., Shanzer, A. & Cahen, D. (2000) Molecular control over Au/GaAs diodes, *Nature* **404,** 166-168.

27. We note that for BSA, an order of magnitude higher currents were measured using the LOFO method, compared with Hg contacts, and this issue is still being studied. One possible explanation is that unlike Az and bR, BSA is not

28. We note here that in our previous work on bR solid-state junctions the observed currents were ~2 orders of magnitudes lower, for a similar junction area (see ref. 16). We attribute this discrepancy to two factors: first, in this previous work the oxide layer was much thicker (oxide grown on Al surface); second, the back contact made now with InGa could not be used on the AlOx surface, and the series resistance then was higher.

29. In principle the high currents could be due partial penetration of the metal contact through the protein monolayers, and we cannot rule this out categorically. However, two facts make this an extremely unlikely scenario: the first is the fact that the results with these two different types of contacts are comparable (in terms of current magnitudes and asymmetry); the second is that to consider tunneling through free space as dominating these junctions, taking $\beta \approx 2$ Å$^{-1}$ for free space, the barrier widths that will result in current levels that were measured with Az, bR and BSA, are 3.3 Å, 3 Å and 5.3 Å, respectively (for the full area of the junction, a case which is not likely in itself if only penetration in between adsorbed molecules is considered).

30. Beratan, D. N., Onuchic, J. N., Winkler, J. R. & Gray, H. B. (1992) Electron-Tunneling Pathways In Proteins, *Science* **258,** 1740-1741.

31. Salomon, A., Cahen, D., Lindsay, S., Tomfohr, J., Engelkes, V. B. & Frisbie, C. D. (2003) Comparison of electronic transport measurements on organic molecules, *Advanced Materials* **15,** 1881-1890.

32. Hopfield, J. J. (1974) Electron-Transfer Between Biological Molecules By Thermally Activated Tunneling, *Proceedings Of The National Academy Of Sciences Of The United States Of America* **71,** 3640-3644.

33. Edwards, P. P., Gray, H. B., Lodge, M. T. J. & Williams, R. J. P. (2008) Electron transfer and electronic conduction through an intervening medium, *Angewandte Chemie-International Edition* **47,** 6758-6765.

34. Paddon-Row, M. N. (2003) Superexchange-mediated charge separation and charge recombination in covalently linked donor-bridge-acceptor systems, *Australian Journal Of Chemistry* **56,** 729-748.

35. McConnell, H. M. (1961) *Journal Of Chemical Physics* **35,** 508.

36. Jortner, J., Bixon, M., Langenbacher, T. & Michel-Beyerle, M. E. (1998) Charge transfer and transport in DNA, *Proceedings Of The National Academy Of Sciences Of The United States Of America* **95,** 12759-12765.

37. Segal, D., Nitzan, A., Davis, W. B., Wasielewski, M. R. & Ratner, M. A. (2000) Electron transfer rates in bridged molecular systems 2. A steady-state analysis of coherent tunneling and thermal transitions, *Journal Of Physical Chemistry B* **104,** 3817-3829.

38. Moser, C. C., Keske, J. M., Warncke, K., Farid, R. S. & Dutton, P. L. (1992) Nature Of Biological Electron-Transfer, *Nature* **355,** 796-802.

39. Kuki, A. & Wolynes, P. G. (1987) Electron-Tunneling Paths In Proteins, *Science* **236,** 1647-1652.

40. Watanabe, J., Morita, T. & Kimura, S. (2005) Effects of dipole moment, linkers, and chromophores at side chains on long-range electron transfer through helical peptides, *Journal Of Physical Chemistry B* **109,** 14416-14425.

41. Malak, R. A., Gao, Z. N., Wishart, J. F. & Isied, S. S. (2004) Long-range electron transfer across peptide bridges: The transition from electron superexchange to hopping, *Journal Of The American Chemical Society* **126,** 13888-13889.

42. Isied, S. S., Ogawa, M. Y. & Wishart, J. F. (1992) Peptide-Mediated Intramolecular Electron-Transfer - Long-Range Distance Dependence, *Chemical Reviews* **92,** 381-394.

43. Felts, A. K., Pollard, W. T. & Friesner, R. A. (1995) Multilevel Redfield Treatment Of Bridge-Mediated Long-Range Electron-Transfer - A Mechanism For Anomalous Distance Dependence, *Journal Of Physical Chemistry* **99,** 2929-2940.

44. Kraatz, H. B., Bediako-Amoa, I., Gyepi-Garbrah, S. H. & Sutherland, T. C. (2004) Electron transfer through H-bonded peptide assemblies, *Journal Of Physical Chemistry B* **108,** 20164-20172.

45. Mujica, V., Ratner, M. A. & Nitzan, A. (2002) Molecular rectification: why is it so rare? *Chemical Physics* **281,** 147-150.

46. Shih, C., Museth, A. K., Abrahamsson, M., Blanco-Rodriguez, A. M., Di Bilio, A. J., Sudhamsu, J., Crane, B. R., Ronayne, K. L., Towrie, M., Vlcek, A., Richards, J. H., Winkler, J. R. & Gray, H. B. (2008) Tryptophan-accelerated electron flow through proteins, *Science* **320,** 1760-1762.

47. Farver, O. & Pecht, I. (1992) Long-Range Intramolecular Electron-Transfer In Azurins, *Journal Of The American Chemical Society* **114,** 5764-5767.

48. Oesterhelt, D. & Stoeckenius, W. (1974) Isolation of the cell membrane of Halobacterium halobium and its fractionation into red and purple membrane, *methods in enzymology* **31,** 667-678.

49. Denkov, N. D., Yoshimura, H., Kouyama, T., Walz, J. & Nagayama, K. (1998) Electron cryomicroscopy of bacteriorhodopsin vesicles: Mechanism of vesicle formation, *Biophysical Journal* **74,** 1409-1420.

50. Ambler, R. P. & Wynn, M. (1973) Amino-Acid Sequences Of Cytochromes C-551 From 3 Species Of Pseudomonas, *Biochemical Journal* **131,** 485-498.